\newcommand{\ud}{\mathrm{d}}

\newcommand{\be}{\begin{equation}}
\newcommand{\beq}{\begin{eqnarray}}
\newcommand{\ee}{\end{equation}}
\newcommand{\eeq}{\end{eqnarray}}

\newcommand{\id}{\mathrm{id}}
\documentclass[twocolumn,10pt]{article}
\makeatletter

\usepackage{graphicx}
\usepackage{color}
\usepackage[normalem]{ulem}
\usepackage{dcolumn}

\usepackage{amsmath}

\usepackage{Luecke}
\usepackage{times}
\usepackage{graphicx}
\usepackage{color}

\newenvironment{sciabstract}{
\begin{quote} \bf}
{\end{quote}}

\newenvironment{scilastnote}{
{\setlength{\labelsep}{.5em}}}

\title{Twin matter waves for interferometry beyond the classical limit} 

\author
{B.~L\"ucke$^1$, M.~Scherer$^1$, J.~Kruse$^1$, L.~Pezz\'e$^2$, F.~Deuretzbacher$^3$, P.~Hyllus$^4$,\\
O.~Topic$^1$, J.~Peise$^1$, W.~Ertmer$^1$, J.~Arlt$^5$, L.~Santos$^3$, A.~Smerzi$^6$, C.~Klempt$^{1\ast}$\\
\\
\normalsize{$^{1}$Institut f\"ur Quantenoptik, Leibniz Universit\"at Hannover, 30167~Hannover, Germany}\\
\normalsize{$^{2}$Laboratoire Charles Fabry, Institut d'Optique, 91127 Palaiseau, France}\\
\normalsize{$^{3}$Institut f\"ur Theoretische Physik, Leibniz Universit\"at Hannover, 30167~Hannover, Germany}\\
\normalsize{$^{4}$Department of Theoretical Physics, The University of the Basque Country, 48080~Bilbao, Spain}\\
\normalsize{$^{5}$QUANTOP, Institut for Fysik og Astronomi, Aarhus Universitet, 8000 \AA{}rhus C, Denmark}\\
\normalsize{$^{6}$INO-CNR BEC Center and Dipartimento di Fisica, Universita` di Trento, 38123 Povo, Italy}\\
\\
\normalsize{$^\ast$To whom correspondence should be addressed; E-mail:  klempt@iqo.uni-hannover.de.}
}

\date{}

\begin{document} 
\maketitle 

\begin{sciabstract}
Interferometers with atomic ensembles constitute an integral part of modern precision metrology. However, these interferometers are fundamentally restricted by the shot noise limit, which can only be overcome by creating quantum entanglement among the atoms. We used spin dynamics in Bose-Einstein condensates to create large ensembles of up to $10^4$ pair-correlated atoms with an interferometric sensitivity $-1.61^{+0.98}_{-1.1}$~dB beyond the shot noise limit. Our proof-of-principle results point the way toward a new generation of atom interferometers.
\end{sciabstract}

Atom interferometers exploit the wave nature of matter, providing a unique tool for modern precision metrology~\cite{Ramsey1950,Borde1989}; most prominently, they have been used to define the second~\cite{Essen1955,Wynands2005}. However, interferometers with uncorrelated atoms are fundamentally limited by shot noise~\cite{Santarelli1999}, a remnant of the atoms' particle nature. This limit can be overcome by creating entanglement between the atoms~\cite{Pezze2009}. Recently, this has been demonstrated for spin-squeezed samples~\cite{Appel2009,Esteve2008,Gross2010,Leroux2010,Riedel2010,Chen2011}. The ultimate Heisenberg limit, where the sensitivity scales linearly with the number of atoms, has been predicted for highly entangled states, such as NOON~\cite{Bollinger1996} and twin Fock states~\cite{Holland1993}.
However, they have only been created with up to five photons~\cite{Kuzmich1998,Ou1999,Walther2004,Mitchell2004,Afek2010} or six ions~\cite{Meyer2001,Leibfried2005}, but not with neutral atoms. Spin dynamics has been proposed as a possible mechanism to turn a large fraction of an atomic Bose-Einstein condensate into a mixture of twin Fock states~\cite{Duan2000,Pu2000}. 

In an interferometric measurement, an ensemble of $N_{\textnormal{tot}}$ particles undergoes a sequence which distributes the particles over two independent modes $+1$ and $-1$. The possible many-particle states are best described by the collective spin $J=N_{\textnormal{tot}}/2$ and the population imbalance $ J_z=( N_{+1}- N_{-1})/2$, where $ N_{\pm 1}$ are the mode populations in the $\pm 1$ mode. The collective spin can be visualized on the generalized Bloch sphere, where the $z$-coordinate represents the population imbalance $J_z$ and the azimuthal angle represents the relative phase between the two modes. The spin uncertainties can be visualized as the extent of the state on the sphere. For instance, unentangled states spin polarized along the x-direction have symmetric uncertainties in relative phase and number, restricted by the uncertainty relation (Fig.~1 A). A typical interferometer sequence results in a rotation of the input state on the Bloch sphere by an angle $\theta$, which depends on the observable of interest. The rotation angle is mapped onto the population imbalance which is finally measured. For unentangled input states, the measurement of the rotation angle is limited by shot noise $\Delta\theta=\sigma(J_z)/(\sqrt{n}\left|d \langle  J_z\rangle /d\theta\right|)\ge 1/\sqrt{n N_{\textnormal{tot}}}$, where $\sigma(J_z)$ is the standard deviation of the population imbalance and $n$ is the number of independent interferometric measurements.

\begin{figure}[ht]
\centering
\includegraphics[width=.90\columnwidth]{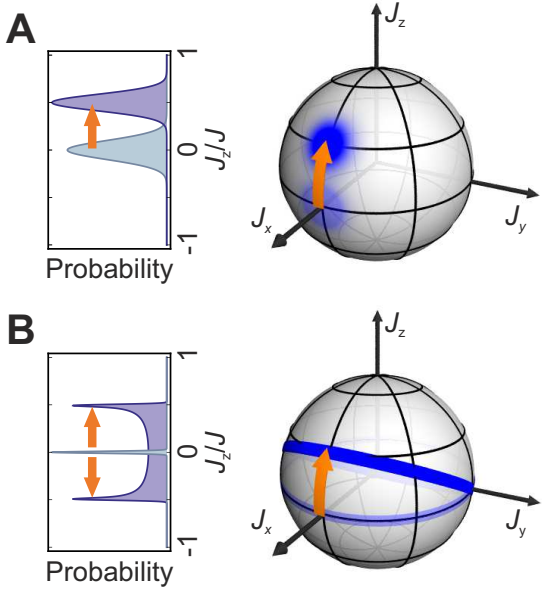}
\caption{Interferometric sensitivity of an unentangled and a twin Fock state. The blue shaded regions represent the uncertainty of the corresponding states. (\textbf{A}) Geometric representation of the sensitivity of an unentangled input state. A rotation on the Bloch sphere is detected by measuring the expectation value of $J_z$. (\textbf{B}) Geometric representation of the sensitivity of a twin Fock input state. Initially, the $\pm 1$ states are equally populated $\left< J_z \right> = 0$ and the relative phase is undefined; the state is represented by a ring around the equator. After a rotation, the $\pm 1$ states are still populated equally $\left< J_z \right> = 0$, but the tilt of the ring indicates fluctuations $\sigma(J_z)$ increasing with the rotation angle.}
\label{fig1}
\end{figure}

A measurement beyond the shot noise limit involves a reduction of the output uncertainty at the expense of an increased uncertainty of the conjugate observable, as has been realized by spin squeezing~\cite{Esteve2008,Gross2010,Riedel2010,Chen2011}. An extreme case is provided by the twin Fock state, where the number of particles in the two modes is exactly equal. Because the relative number uncertainty vanishes, the relative phase is completely undetermined. The twin Fock state is hence represented by an arbitrarily thin line around the equator of the Bloch sphere (Fig.~1 B). A rotation of this state maps the quantity of interest not on the expectation value of the output $\langle J_z\rangle$ but rather on its fluctuations $\langle J_z^2\rangle$~\cite{Kim1998}. The resulting sensitivity overcomes the shot noise limit, and is only bound by the Heisenberg limit $\Delta\theta \propto 1/N_{\textnormal{tot}}$. Such an interferometer with twin Fock input states has been created for very small samples, including single~\cite{Kuzmich1998} and double~\cite{Ou1999,Nagata2007} pairs of photons, and a single pair of $^9$Be$^+$ ions~\cite{Meyer2001}.

We use spin dynamics in Bose-Einstein condensates for the creation of up to $10^4$ paired neutral atoms in two different spin states. To generate these nonclassical states of matter, we start with a Bose-Einstein condensate with horizontal spin orientation (Zeeman substate $m_F=0$). In these ensembles, collisions may produce correlated pairs of atoms with spins up and down~($m_F=\pm 1$)~\cite{Duan2000,Pu2000}. These collisions are bosonically enhanced if the output modes are occupied. Therefore, they act as a parametric amplifier for a finite initial population in $m_F=\pm 1$ or for pure vacuum fluctuations~\cite{Klempt2010}. During the parametric amplification of vacuum~\cite{Barnett1990}, the total number of atoms produced in $m_F=\pm 1$ and its fluctuations increase exponentially with time. The conjugate variable of the total number is the sum of the two atomic phases, whose fluctuations are exponentially damped~\cite{Scherer2010}. Furthermore, the number difference between $m_F=\pm 1$ atoms is zero (without fluctuations), and hence the corresponding conjugate variable, the relative phase, is fully undetermined. The underlying physics closely resembles that of optical parametric down-conversion in nonlinear crystals, currently the most important technique to generate nonclassical states of light.

The experiments are started by creating a $^{87}$Rb condensate of $2.8 \times 10^4$ atoms in the hyperfine state $F=2$, $m_F=0$ in an optical dipole trap (Fig.~2 A). We initiate the spin dynamics at a magnetic field of $1.23$~G, where an excited spatial mode is populated and vacuum fluctuations are amplified~\cite{SOM}. The states $F=2$, $m_F=\pm1$ are populated by spin dynamics for an optimal duration of $15$~ms~\cite{SOM}. Subsequently, the dipole trap is switched off and all three spin components are recorded by absorption imaging (Fig.~2 B,C).

\begin{figure*}[htb]
\centering
\includegraphics[width=1.9\columnwidth]{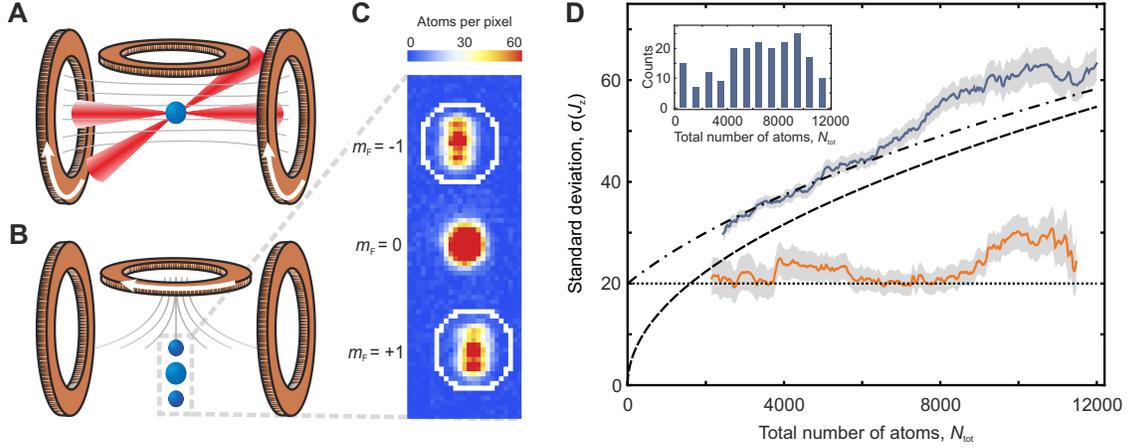}
\caption{ Preparation and analysis of the input state. (\textbf{A}) At a homogeneous magnetic field, a Bose-Einstein condensate in a crossed-beam optical dipole trap generates atoms in the states $m_F=\pm1$. (\textbf{B}) Afterwards, the trap is switched off and the three Zeeman sublevels are split by a strong magnetic field gradient. (\textbf{C}) Finally, the three clouds are detected by absorption imaging and the number of atoms is counted within circular masks (white line). (\textbf{D}) Distribution of the relative number of atoms in $m_F=\pm 1$. The measured standard deviation of the population imbalance $\sigma (J_z)$ (orange line) is well below the shot noise limit (dashed line) and mainly limited by the detection noise (dotted line). The experimental result for unentangled atoms (blue line) corresponds to the combination of shot noise and detection noise (dash-dotted line). The shaded area indicates the error of the standard deviation. Inset: Distribution of the total number of atoms. The total number of atoms $N_{\textnormal{tot}}$ fluctuates strongly as a result of the amplification of vacuum fluctuations.}
\label{fig2}
\end{figure*}

A test of this sensitivity requires the implementation of an internal-state beam splitter. To drive the transition connecting the $F=2$, $m_F=\pm 1$ states, we employ three resonant microwave pulses (Fig.~3 A). On the Bloch sphere, the full sequence represents a rotation around the x-axis by an angle $\theta=\tau\; \Omega_{\rm R}$, where $\tau$ represents the duration of the coupling pulse and $\Omega_{\rm R}$ is the Rabi frequency. Figure~3 C shows the standard deviation of the normalized population imbalance $\sigma(J_z/J)$ as a function of the rotation angle $\theta$. It illustrates that the phase to be measured in the interferometer is mapped onto the fluctuations of the output instead of the expectation value. As the picture of the rotating ring on the Bloch sphere suggests, the standard deviation oscillates approximately as $\sigma(J_z/J)=\alpha \left|\textrm{sin}\;\theta\right|$, where $\alpha$ characterizes the interferometric contrast. Similar to optical interferometers, the contrast relies heavily on an identical spatial mode of the two input states. Although single-mode operation can be guaranteed for small samples in tight traps, the creation of large samples demands more careful consideration. Generally, spin dynamics populates various spatial modes~\cite{Klempt2009}, creating multiple twin Fock states without a common phase relation. After the beam splitter, a multi-mode state yields a number distribution which is a convolution of many single-mode distributions - resulting in a severe reduction of the contrast. We have optimized the configuration for single-mode operation~\cite{SOM} and reach a contrast $\alpha=0.67(0.04)$, close to the ideal value of $\alpha=1/\sqrt{2}\approx 0.71$.

\begin{figure*}[!ht]
\centering
\includegraphics[width=1.9\columnwidth]{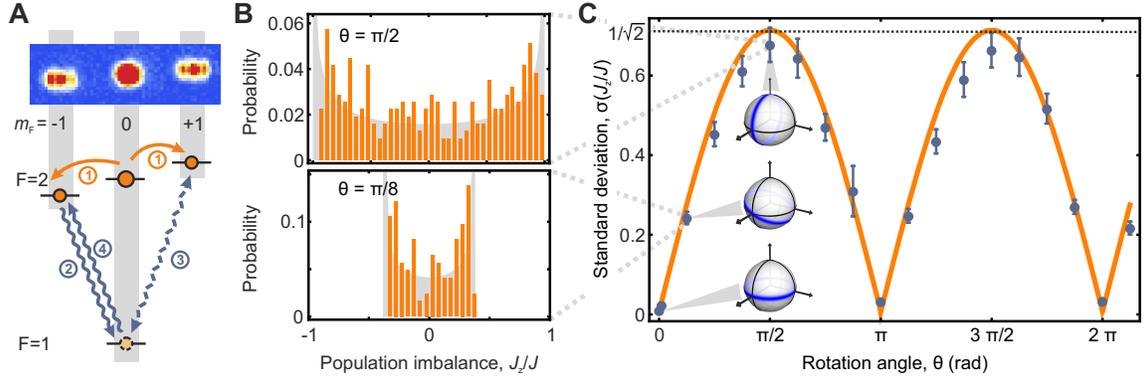}
\caption{ Internal-state beam splitter for nonclassical matter waves. (\textbf{A}) Schematic of the beam splitter sequence. (1) Spin dynamics initially populates the states $|F=2,m_{ F} = \pm 1\rangle$. (2) To couple these states, the atoms in $|2,-1\rangle$ are transferred to $|1,0\rangle$ by a microwave pulse. (3) Next, a pulse of variable duration $\tau$ couples the states $|2,+1\rangle$ and $|1,0\rangle$. (4) Finally, atoms in $|1,0\rangle$ are transferred to the state $|2,-1\rangle$ to enable their independent detection. (\textbf{B}) Distribution of the normalized population imbalance for two coupling pulse durations. The strongest fluctuations are obtained for a coupling pulse duration corresponding to a symmetric $\theta=\pi/2$ beam splitter. The shaded area corresponds to the ideal result. (\textbf{C}) Fluctuation of the normalized population imbalance. As expected from the representation on the Bloch sphere, the standard deviation $\sigma (J_z / J)$ oscillates as $\left| \sin \theta \right| / \sqrt{2}$  as a function of the rotation angle $\theta$. All quantities were obtained for $N_{\textnormal{tot}}$ between $3,000$ and $8,000$ atoms.}
\label{fig3}
\end{figure*}

The symmetric beam splitter obtained for $\theta=\pi/2$ is of special interest. It turns the ring on the Bloch sphere onto the $xz$-plane, leading to a maximal uncertainty of the population imbalance, and consequently a minimal relative phase uncertainty. It is therefore an optimal state to detect the phase evolution during a Ramsey sequence. Figure~3 B shows the distribution of the normalized population imbalance. For $\theta=\pi/2$, it assumes a characteristic shape~\cite{Campos1989} which reflects a high probability of detecting most of the atoms in one of the output ports. Ideally, either all even or all odd numbers should appear in the distribution of the population imbalance $J_z$, but this parity effect is not accessible within our experiments due to detection noise and particle losses. 

Based on the beam splitter, we demonstrate that the experimentally created state is entangled and useful for sub-shot-noise interferometry. The phase estimation uncertainty $\Delta\theta$ is inferred from the state's sensitivity to small rotations around an arbitrary axis in the x-y plane. It is deduced from error propagation according to $\Delta\theta = \Delta J_z^2 / (\sqrt{n}\left|d \left\langle J_z^2\right\rangle / d\theta \right|)$ for $n$ independent repetitions of the interferometric measurement. In our case, we only rotate the state around the x-axis, which is less susceptible to technical noise than a Ramsey sequence due to its short duration. The expectation values $\left\langle J_z^2\right\rangle$ and $\left\langle(\Delta J_z^2)^2\right\rangle = \left\langle J_z^4\right\rangle - \left\langle J_z^2\right\rangle^2$ are shown in Figs.~4 A,B for small rotation angles $\theta$. The result agrees well with the ideal case including a finite detection noise~\cite{SOM}. The slight deviation for larger angles indicates additional technical noise, such as magnetic field noise or microwave intensity noise. Polynomial fits~\cite{SOM} allow for a precise estimate of the sensitivity (Fig.~4 C) according to the error propagation. At the optimal point ($\theta=0.015$), we reach a measurement uncertainty $\Delta\theta/\Delta\theta_{\textnormal{sn}} = 0.83(0.1)$, which is $-1.61^{+0.98}_{-1.1}$~dB below the shot noise limit $\Delta\theta_{\textnormal{sn}} =1/\sqrt{n \langle N_{\textnormal{tot}} \rangle}$. This result is also $-2.5^{+0.98}_{-1.1}$~dB below the optimal classical result achievable with our apparatus when both shot noise and detection noise are considered.
\begin{figure*}[!htb]
\centering
\includegraphics[width=1.2 \columnwidth]{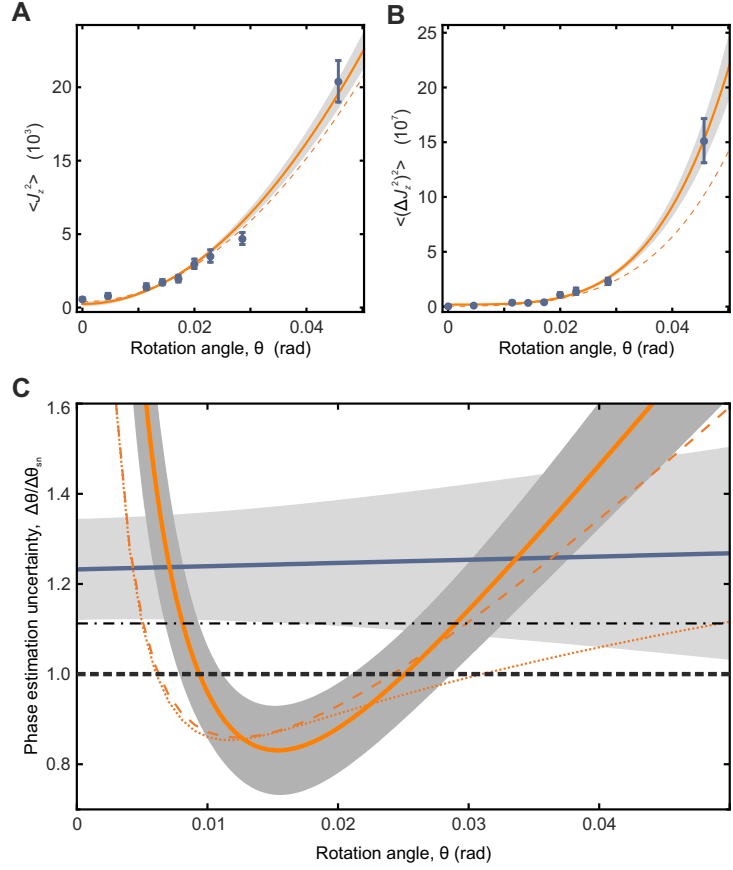}
\caption{ Phase sensitivity of the pair-correlated state. (\textbf{A}) Variance of the population imbalance $\langle \sigma(J_z)^2\rangle=\langle  J_z^2\rangle$ for a total number $N_{\textnormal{tot}}$ between $6,400$ and $7,600$ atoms as a function of the rotation angle $\theta$. The orange line is a quadratic fit to the data~\cite{SOM} and the shaded area indicates the uncertainty of the fit. Orange dashed line: the theoretical prediction for a detection noise of $20$~atoms. (\textbf{B}) Same for the fourth moment of the population imbalance $\langle(\Delta  J_z^2)^2\rangle=\langle  J_z^4\rangle-\langle  J_z^2\rangle^{2}$. (\textbf{C}) Sensitivity of the nonclassical state. The phase estimation uncertainty $\Delta\theta$ with respect to shot noise $\Delta\theta_{\textnormal{sn}}=1/\sqrt{n \langle N_{\textnormal{tot}}\rangle}$ (solid orange line) is obtained from the fits in (A) and (B) and compared to the expected sensitivity (dashed orange line). At $\theta \approx 0.015$~rad the phase sensitivity lies $-1.61^{+0.98}_{-1.1}$~dB below the shot noise limit (black dashed line). For a range of values of $\theta$ it is very close to the best achievable value given by the Cramer-Rao lower bound (dotted orange line)~\cite{SOM}. The experimentally measured sensitivity of an unentangled state (solid blue line) is measured by applying a $\pi/2$ beam splitter to a spin polarized condensate. The technical noise of the beam splitter causes the measured sensitivity to lie slightly above the combination of shot noise and detection noise (black dash-dotted line).}
\label{fig4}
\end{figure*}

The signature of useful entanglement for sub-shot-noise interferometry is a Fisher information larger than the total number of particles $\mathcal{F}> N_{\textnormal{tot}}$~\cite{Pezze2009}. The Fisher information also provides the ultimate achievable sensitivity via the Cramer-Rao bound $\Delta\theta\ge1/\sqrt{n\mathcal{F}}$, which can be reached asymptotically for a large number of measurements $n$. Notice, however, that here, as well as in previous experiments on squeezing, entanglement and quantum interferometry with ultracold atoms, the total number of particles fluctuates in different realizations; thus, the entanglement criteria must be generalized by replacing the fixed number of particles $N_{\textnormal{tot}}$ with its statistical average $\langle N_{\textnormal{tot}}\rangle$ \cite{Hyllus2010}. A lower bound for the Fisher information is provided by $\mathcal{F}\ge \left|d \left\langle J_z^2\right\rangle / d\theta \right|^2 / (\Delta J_z^2)^2$. In our experiment, this lower bound is maximized for an angle of $0.015$~rad, resulting from the finite detection efficiency. At this angle, we obtain $\mathcal{F}/\langle N_{\textnormal{tot}}\rangle \geq 1.45^{+0.42}_{-0.29}$, where a value greater than 1 proves entanglement. This entanglement can be exploited for sub-shot-noise phase estimation in a large variety of interferometer configurations~\cite{hyllusPRA10}, including the Ramsey interferometer. 

The presented results are mainly restricted by the detection limit of 20~atoms. A detection limit of $5$ atoms would allow for a minimum phase estimation uncertainty of $-13.6$~dB below the shot noise limit~\cite{SOM}. Furthermore, the region of sub-shot-noise interferometry can be extended up to the whole phase domain by employing Maximum Likelihood or Bayesian phase estimation protocols. The combination of large samples and quantum enhanced sensitivity should open exciting perspectives for highly sensitive measurements with a new generation of atom interferometers.

Bücker et al. \cite{Bucker2011} recently reported reduced atom number fluctuations in twin-atom beams, and the group of M. Oberthaler has independently used spin dynamics in Bose-Einstein condensates to generate atomic two-mode entanglement detected by a homodyning technique.

\begin{scilastnote}
\textbf{Acknowledgments.} We thank E. Rasel for stimulating discussions and P. Zoller for helpful remarks concerning the manuscript. We acknowledge support from the Centre for Quantum Engineering and Space-Time Research QUEST, the European Science Foundation (EuroQUASAR) and the Danish National Research Foundation Center for Quantum Optics. P.H. acknowledges financial support of the ERC Starting Grant GEDENTQOPT.
\end{scilastnote}

%% Supporting Online Material
\newpage
\clearpage
\setcounter{figure}{0}

\renewcommand{\thefigure}{S~\arabic{figure}}			% figure enumeration with "S"
\renewcommand{\theequation}{S~\arabic{equation}}					 % Equation counter with "S"

\section*{Supporting Online Material}

\section{Magnetic field dependent spin resonances}
\label{Magnetic_dependence}

\begin{figure}[!htb]
\centering
\includegraphics[width=.9\columnwidth]{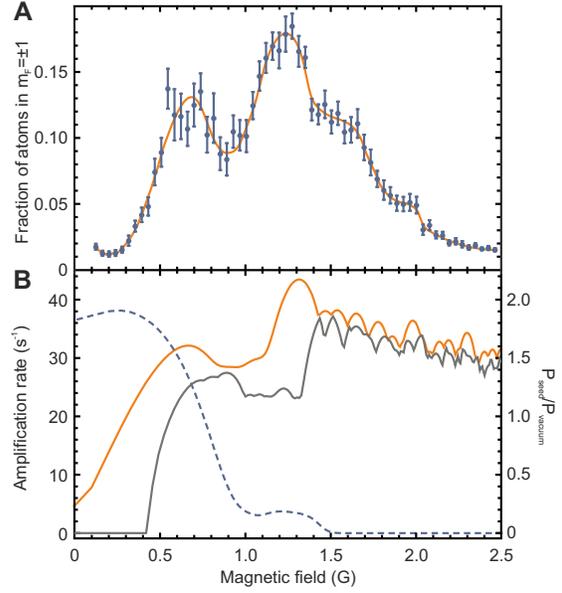}
\caption[Figure S1]{Magnetic field dependence of the spin dynamics. (\textbf{A}) Fraction of atoms transferred to the $m_F= \pm 1$ states versus the magnetic field after an evolution time of 18~ms. A first resonance is observed at 0.7 G and a second one at 1.23 G. (\textbf{B}) Numerical calculation of the amplification rate of the two most unstable excitation modes (solid lines, left axis). The positions of the maxima reflect the experimentally observed resonances. The dashed line (right axis) shows the ratio between the population triggered by vacuum fluctuations $\text{P}_{\textnormal{vacuum}}$ and by residual seed atoms $\text{P}_{\textnormal{seed}}$ at 18~ms. The calculation indicates the almost pure quantum character of the second resonance.}
\label{SIfig1}
\end{figure}

The amplification of vacuum fluctuations is only possible for certain magnetic fields. Figure~\ref{SIfig1} A shows the fraction of atoms transferred to the state $m_F= \pm 1$ as a function of the applied magnetic field, after an evolution time of $18$~ms. The resonance structure reflects the spin dynamics rate of various spatial modes~(30). This can be visualized in a simplified model, in which the effective trap experienced by the $m_F=\pm 1$ atoms is substituted by a cylindrical box potential~(28). The corresponding eigenmodes resemble the TEM laser modes. Each spatial mode results in a spin dynamics resonance, if its eigenenergy matches the quadratic Zeeman energy of the specific magnetic field. We attribute the strongest resonance at $1.23$~G to a spatial mode similar to a Hermite-Gaussian TEM$_{\text{20}}$ laser mode~(28), apparent in the density profile of the transferred atoms.

The magnetic field resonances are precisely described by means of a Bogoliubov-de Gennes theory~(30). The presence of unstable spin modes leads to an exponential amplification of the population of the $m_F=\pm 1$ states, which depends on the magnetic field (Fig.~\ref{SIfig1} B, solid lines). This amplification can be triggered both by a residual number of atoms in $m_F= \pm 1$ (seed atoms) and by pure vacuum fluctuations~(26). The dashed line displays the calculated ratio between the populations triggered by these two mechanisms. Whereas the spin dynamics at low magnetic fields is dominantly triggered by seed atoms, the reduced mode overlap leads to a vanishing influence of the seed at high magnetic fields. In our experiments, the spin dynamics is initiated at a magnetic field of $1.23$~G. This resonance provides the strongest transfer rate and a negligible contribution of seed atoms. Thus it ensures an efficient amplification of almost pure vacuum fluctuations. Note the large separation between the two most unstable modes at 1.23~G, which provides single-mode operation as discussed in Sec.~\ref{Single_mode_operation}.

\section{Experimental sequence and imaging}

The experimental sequence starts by preparing an ultracold thermal cloud of $^{87}$Rb in the state $F=2$, $m_F=2$ in an optical dipole trap. The dipole trap is generated by two crossed laser beams at a wavelength of $1064$~nm. The atoms are transferred to the state $F=1$, $m_F=1$ by a resonant microwave $\pi$ pulse to avoid hyperfine-changing collisions during the subsequent evaporation to Bose-Einstein condensation (BEC). A second microwave pulse of variable duration transfers an adjustable number of atoms to the state $F=2$, $m_F=0$. To provide a well-defined starting time for the spin dynamics and to avoid spurious atoms, a third microwave pulse transfers the atoms to the state $F=1$, $m_F=0$ and a strong magnetic field gradient removes all residual atoms in magnetic field sensitive states. At the same time, the dipole trap is adjusted to its final trapping frequencies of $2 \pi \times (183, 151, 122)$~Hz. Spin dynamics is initiated by a final transfer to the state $F=2$, $m_F=0$. During the evolution time, a constant magnetic field is actively stabilized to $1.23$~G (with shot-to-shot fluctuations below $0.5$~mG). The chosen magnetic field allows for the controlled population of a single spatial mode, where vacuum fluctuations are amplified (Section~\ref{Magnetic_dependence}). For some of the experiments, the two states $F=2$, $m_F=\pm1$ are coupled at the end of the evolution time by a series of microwave pulses as shown in Fig.~3 A in the main text.

\begin{figure}[!ht]
\centering
\includegraphics[width=.8\columnwidth]{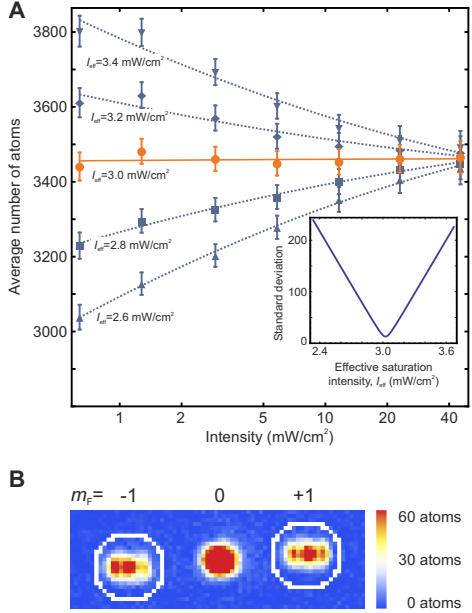}
\caption{Measurement of the number of atoms. (\textbf{A}) The effective saturation intensity is determined by preparing atoms in the $m_{F}=\pm 1$ states using a radio frequency pulse. These ensembles are detected with different imaging beam intensities. For the correct effective saturation intensity, the number of atoms does not depend on the imaging beam intensity~\cite{Reinaudi2007}. Minimizing the standard deviation of the mean number of atoms (inset) yields $3.02(6)$~{mW/cm}$^2$.
(\textbf{B}) Typical density profile of the atomic clouds. Disk-shaped masks (white) are centred around the clouds of atoms in the $m_F=-1$ (left) and $m_F=+1$ (right) states. The number of atoms within each mask is summed to calculate the population of each cloud. The size of the masks is carefully chosen to detect the complete atom clouds and to minimize the detection noise. The effect of averaging over single pixels is checked to be below $1~\%$.} 
\label{SIfig:2}
\end{figure}

Finally, the dipole trap is switched off to allow for self-similar expansion. During a time of flight of $5$~ms, the different Zeeman states are split by a strong magnetic field gradient. At a final magnetic field of $3.6$~G along the imaging direction, the atomic density is recorded by resonant absorption imaging on the $F=2 \rightarrow F'=3$ cycling transition. A pulse of circularly polarized light with a duration of $70~\mu$s and an intensity of $1.6$~mW/cm$^2$ is imaged onto a CCD camera. The atomic column density is inferred from these pictures by employing an effective saturation intensity. This effective saturation intensity of $3.02(6)$~mW/cm$^2$ is determined according to Ref.~\cite{Reinaudi2007} as shown in Fig.~\ref{SIfig:2} A. The accuracy of the employed method was checked in Ref.~9. The total number of atoms in each cloud is calculated by a summation of all single-pixel atomic densities within a disk-shaped mask (Fig.~\ref{SIfig:2} B).

\section{Time evolution of spin dynamics}\label{time_evolution}

Only for short evolution times, spin dynamics leads to the undisturbed parametric amplification of vacuum fluctuations. However, the creation of large samples requires a sufficiently long evolution time. It is hence important to investigate the evolution time to deduce an optimal duration. 

\begin{figure}[!ht]
\centering
\includegraphics[width=.9\columnwidth]{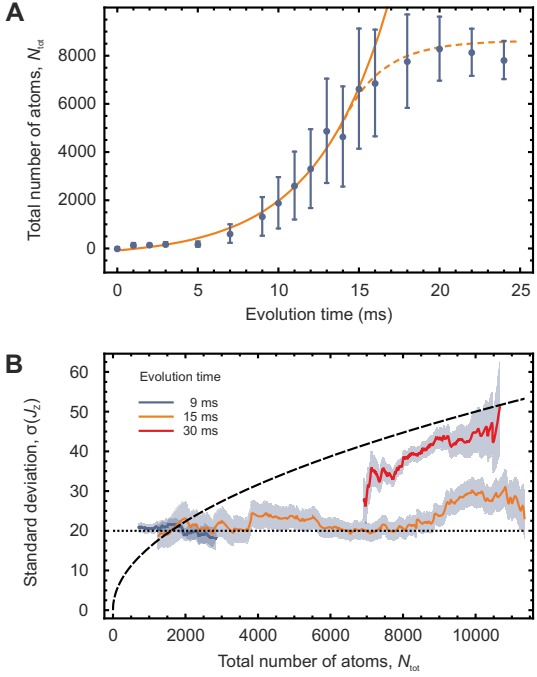}
\caption{Time evolution of spin dynamics. (\textbf{A}) Total number of atoms in the $m_F=\pm $1 states and its standard deviation versus evolution time. The number of transferred atoms (circles) increase exponentially (solid line) depending on the evolution time. After 15~ms a saturation sets in (dashed line). (\textbf{B}) Standard deviation of the population imbalance $\sigma(J_z)$ for different evolution times. After 9~ms only a small number of atoms is transferred to the $m_F=\pm $1 states, whereas after an evolution time of $15$~ms the number of atoms is strongly increased. For both times the standard deviation is dominated by the detection noise (dotted line). After an evolution time of 30~ms, the fluctuations of the populations of $m_F=+1$ and $m_F=-1$ increase and approach the shot noise limit (dashed line).  The standard deviation for a given total number of atoms is calculated by taking into account an interval of $\pm 1000$ atoms. The grey shaded area indicates the error of the standard deviation. }
\label{SIfig:3}
\end{figure}

The exponential amplification of the total number of atoms in the $m_F=\pm 1$ states is shown in Fig. \ref{SIfig:3} A. The mean number increases exponentially up to an evolution time of $15$~ms. Within this time, the standard deviation of the total number also increases exponentially, as predicted for the amplification of vacuum~(27). For longer evolution time, however, four effects lead to a saturation of the total number. Firstly, the number of atoms in the $m_F=0$ condensate, initially $28,000$ atoms, is depleted, which reduces the production of $m_F=\pm1$ atoms. Furthermore, the density of $m_F=\pm1$ atoms is not negligible and leads to a transfer back to the $m_F=0$ state. This density also leads to a population of the $m_F=\pm 2$ states. Finally, hyperfine-changing collisions result in a steady loss of atoms, even though the lifetime is large compared to the typical evolution times considered. These effects scale with the total number of atoms in $m_F=\pm1$ and therefore also reduce the standard deviation for long evolution time. 

These processes also affect the fluctuation of the population imbalance. Figure \ref{SIfig:3} B shows the standard deviation of the population imbalance $\sigma(J_z)$ as a function of the total number of atoms $N_{\text{tot}}$ for three different evolution times. It is compared to the detection noise of $20(1)$ atoms which we have independently measured by recording 400 pictures without spin dynamics. Note that we define the detection noise in terms of $J_z=(N_{+1}-N_{-1})/2$. Both for $9$~ms and $15$~ms, the standard deviation is comparable to the detection limit and shows almost no dependence on the total number. For longer evolution times, represented here by $30$~ms, the fluctuations start to rise and develop a dependence on the total number of atoms. These measurements result in an optimal evolution time of $15$~ms, where a substantial population of the $m_F=\pm 1$ states is reached and the fluctuations of the population imbalance are much smaller than the detection noise.

\section{Single-mode operation}\label{Single_mode_operation}

\begin{figure}[!ht] 
\centering
\includegraphics[width=.9\columnwidth]{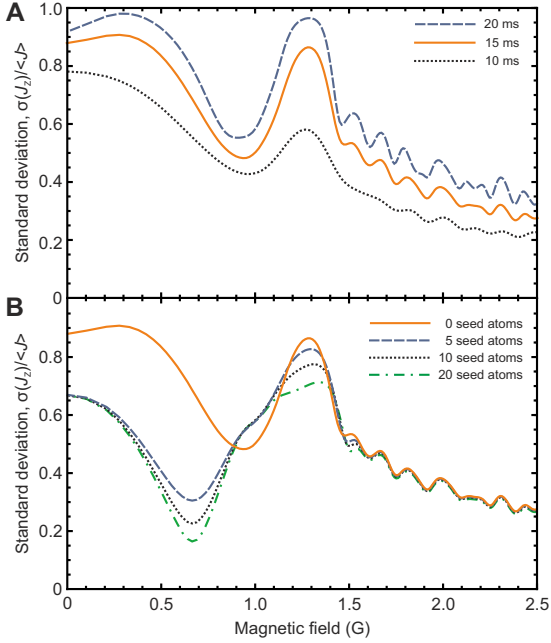}
\caption{Fluctuations of the population imbalance after the beam splitter. 
(\textbf{A}) Numerical results for the magnetic field dependence of the fluctuation of the population imbalance $\sigma(J_z)/\langle J \rangle$ after applying a rotation angle $\pi/2$
for different evolution times. The fluctuations are maximal close to the spin dynamics resonances and present a maximmum at $1.3$~G. (\textbf{B}) Fluctuations of the difference of the number of atoms for an evolution time of $15$ms for various numbers
of spurious seed atoms. At low magnetic fields fluctuations are strongly
reduced by seed atoms. On the contrary, since the second resonance at higher
magnetic field is mainly triggered by vacuum fluctuations, the
seed dependence is strongly reduced. Note that contrary to Fig.~3 C in the main text, the optimal standard deviation is one.}
\label{SIFig4}
\end{figure}

The $m_F=\pm 1$ atoms may populate multiple spatial modes as discussed in Section \ref{Magnetic_dependence}. This results in a magnetic field dependence of the standard deviation of the population imbalance after the beam splitter (Fig.~\ref{SIFig4} A). For an optimal interferometric sensitivity, it is necessary to maximize this standard deviation.

Insight may be gained from the simplified model using a cylindrical box potential presented in Section~\ref{Magnetic_dependence}. This model shows that an optimal standard deviation is achieved when the most unstable spin mode is nondegenerate. When this is not the case, the standard deviation is reduced by the square root of the number of unstable degenerate modes contributing to the spin dynamics, since the ensembles in these modes are independent. This value is further reduced by up to $1/\sqrt{2}$ when spurious seed atoms trigger the dynamics (Fig. \ref{SIFig4} B). Hence, an optimal configuration allows for the amplification of a single nondegenerate mode, which is significantly more unstable than the rest, and has a low sensitivity to spurious seeding.

In our experiment, the spacing between low-energy modes of the effective trap is slightly smaller than the typical energy for spin-changing collisions. This leads to an unavoidable population of various modes, which in turn reduces the standard deviation of the population imbalance. However, during the amplification, the relative importance of additional modes decreases exponentially compared to that of the dominant one, i.e. the system approaches the ideal single-mode operation with increasing time. 

A full numerical multi-mode calculation employing realistic experimental parameters shows that the resonance belonging to the nondegenerate TEM$_{20}$ mode is most favourable. As expected, the standard deviation grows with the evolution time (Fig.~\ref{SIFig4} A). However, this time cannot be arbitrarily long as discussed in section \ref{time_evolution}. The best compromise is achieved for an evolution time of $15$~ms. Note that although the best value of the variance is achieved at very low magnetic fields around $0.4$~G, it is unfavourable to operate at those low magnetic fields, mainly due to spurious seeding.

\section{Calculation of the interferometric sensitivity}

In the beam-splitting process the experimentally created state is transformed by the collective spin rotation $e^{-i \theta  J_x}$~\cite{Yurke1986}. For a fixed value of $\theta$, the experiment provides a sequence of independent results, $N_{+1}$ and $N_{-1}$, for the number of particles in the $m_F=+1$ and $m_F=-1$ states, respectively. For the analysis of the phase uncertainty, we select the data in the interval $N_{\mathrm{tot}} = N_{+1} + N_{-1} = 7,000 \pm 600$ atoms. This is chosen for two reasons: First, shot noise is large with respect to the detection noise $\sqrt{N_\textnormal{tot}} \gg 2 \sigma_\text{dn}$, where $\sigma_\text{dn} \sim 20$ is our detection noise. Second, this interval is close to the maximum of the distribution of the total number of particles (Fig.~2 D) such that the chosen interval contains a relatively large number of data points ($p\sim 100$), for each value of $\theta$. By using the selected data, we evaluate the expectation value of $ J_z^\alpha$ as the classical average: $\langle  J_z^\alpha \rangle = \sum_{i=1}^p \big(N_R^{(i)}\big)^\alpha/p$, where $N_R^{(i)} = \big( N_{+1}^{(i)}-N_{-1}^{(i)} \big)/2$ with $i=1,...,p$ and $\alpha$ is an integer number. The values of $\langle  J_z^2 \rangle$ and $(\Delta  J_z^2)^2 = \langle  J_z^4\rangle-\langle  J_z^2\rangle^2$ obtained experimentally are shown in Figs.~4~A and 4~B for 9 values of the rotation angle $\theta$. The corresponding mean square fluctuations $\sigma^2_{\langle  J_z^2 \rangle}$ and $\sigma^2_{(\Delta  J_z^2)^2}$, shown as error bars in Figs.~4~A and 4~B, are calculated from the experimental data and hence take the finite number of measurements and the experimental classical noise into account.

For a general distribution of twin Fock states, $\rho_{\mathrm{inp}} = \sum_N p_N \vert N/2, N/2 \rangle \langle N/2, N/2 \vert$ (where $p_N$ are arbitrary coefficients with $p_N>0$, $\sum_N p_N=1$), and assuming a Gaussian detection noise $\sigma_\text{dn}$ \cite{Eckert2006}, we find 
\be \label{gJz2}
\langle  J_z^2 \rangle_{\mathrm{gdn}} = \langle  J_z^2 \rangle_{\id} + \sigma_\text{dn}^2
\ee  
and 
\be \label{gDJz2}
(\Delta  J_z^2)^2_{\mathrm{gdn}} = (\Delta  J_z^2)^2_{\id} + 4\sigma_\text{dn}^2 \langle  J_z^2 \rangle_{\id} + 2 \sigma_\text{dn}^4,
\ee  
where the expectation values are calculated for the output state $\rho_{\text{out}} = e^{- i \theta  J_x} \rho_{\text{inp}} e^{+ i \theta  J_x}$. The ideal expectation values in these expressions are 
\be \label{idJz2} 
\langle  J_z^2 \rangle_{\id} = \frac{ \langle N_{\mathrm{tot}}^2 \rangle + 
2  \langle N_{\mathrm{tot}} \rangle}{8} \sin^2 \theta
\ee
and 
\be \label{idDJz2}
%\begin{split}
\textstyle(\Delta  J_z^2)^2_{\id} = \bigg( \frac{\langle N_{\mathrm{tot}}^4 \rangle}{128} 
+ \frac{\langle N_{\mathrm{tot}}^3 \rangle}{32} 
- \frac{\langle N_{\mathrm{tot}}^2 \rangle}{32} 
- \frac{3\langle N_{\mathrm{tot}} \rangle}{8} \bigg) \sin^4 \theta
+ \\ \langle  J_z^2 \rangle_{\id},
%\end{split}
\ee
being $\langle N_{\mathrm{tot}}^\alpha \rangle = \sum_N p_N \, N^\alpha$. Equations (\ref{gJz2}) and (\ref{gDJz2}) are shown as dashed orange lines in Figs.~4~A and 4~B. 

\begin{figure}[!hbt]
\begin{center}
\includegraphics[width=.9\columnwidth]{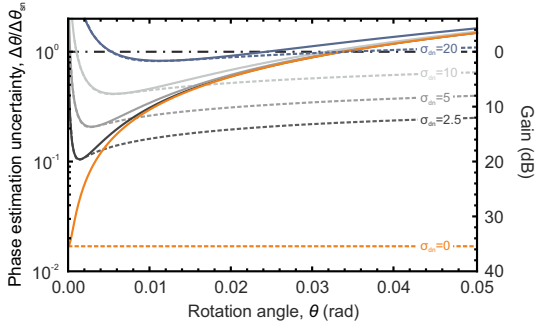} 
\end{center}
\caption{Expected phase estimation uncertainty.
Phase estimation uncertainty with respect to shot noise calculated for a twin Fock input state $\vert \psi \rangle_{\mathrm{inp}}=\vert N_{\mathrm{tot}}/2, N_{\mathrm{tot}}/2 \rangle$ with $N_{\mathrm{tot}}=7000$, and a Gaussian model of detection noise. The phase estimation uncertainty is calculated with Eq.~(\ref{ErrProp}) (solid lines) and the Cramer-Rao bound Eq.~(\ref{CR}) (dashed lines). Different colors refer to different values of detection noise $\sigma_\text{dn}$. $\sigma_\text{dn}=20$ corresponds to the case of our experiment (blue lines). The horizontal dot-dashed line is the shot-noise limit.}
\label{FigSup}
\end{figure}

The phase estimation uncertainty is obtained from the error propagation equation
\be \label{ErrProp}
\Delta \theta = \frac{\Delta  J_z^2}{ \sqrt{n} \vert \ud \langle  J_z^2 \rangle/\ud \theta \vert}.
\ee
In the ideal case, by using Eqs.~(\ref{idJz2}) and (\ref{idDJz2}), the phase estimation uncertainty strongly depends on the phase shift. Its optimal value, $\Delta \theta = \sqrt{2}/\sqrt{n}\sqrt{\langle N_{\mathrm{tot}}^2 \rangle + 2 \langle N_{\mathrm{tot}} \rangle}$, is reached at $\theta=0$. As shown in Fig.~\ref{FigSup}, the phase estimation uncertainty rapidly decreases for larger values of $\theta$ and becomes worse than the shot noise.

For the Gaussian noise model discussed above, Eq.~(\ref{ErrProp}) is calculated by using Eqs.~(\ref{gJz2}) and (\ref{gDJz2}). For small values of $\theta$, the distribution of the population imbalance is dominated by the detection noise: The derivative $\ud \langle  J_z^2 \rangle_{\mathrm{gdn}}/\ud \theta$ vanishes, while $(\Delta  J_z^2)^2	_{\mathrm{gdn}}$ remains finite. The phase estimation uncertainty predicted by Eq.~(\ref{ErrProp}) thus diverges at $\theta=0$. The minimum of Eq.~(\ref{ErrProp}) is obtained approximately when $\langle  J_z^2 \rangle_{\mathrm{id}} \sim \sigma_\text{dn}^2$. A plot of the phase estimation uncertainty for different values of $\sigma_\text{dn}$ is shown 
in Fig.~\ref{FigSup}. 

To calculate the phase estimation uncertainty from the experimental results, we take error bars on the measured average moments into account. For each $\theta$, we assume that we are in the central limit and apply a parametric bootstrap method~\cite{Efron1994}: For each experimental value of $\theta$, we generate a pair of random numbers, 
the first (second) normally distributed with mean $\langle  J_z^2 \rangle$ [$(\Delta  J_z^2 )^2$] and variance $\sigma^2_{\langle  J_z^2 \rangle}$ [$\sigma^2_{(\Delta  J_z^2 )^2}$]. These values are then fitted according to the function 
$f_{\langle  J_z^2 \rangle}(\theta) = a_1 \sin^2 \theta + b_1$ 
[$f_{(\Delta  J_z^2)^2}(\theta) = a_2 \sin^4 \theta + b_2 \sin^2\theta + c_2$], 
where $a_1$ and $b_1$ [$a_2$, $b_2$ and $c_2$] are fitting parameters. The choice of the fitting functions follows from Eqs.~(\ref{gJz2}) to (\ref{idDJz2}). The phase uncertainty is then calculated by using Eq.~(\ref{ErrProp}), 
i.e. $(\Delta \theta)^2 = f_{(\Delta  J_z^2)^2}(\theta)
/n\vert \ud f_{\langle  J_z^2 \rangle}(\theta)/\ud \theta \vert^2$.
This procedure is repeated several times to guarantee statistical significance. The average fitting functions for $\langle  J_z^2 \rangle$ and $(\Delta  J_z^2 )^2$ are shown as solid orange lines in Figs.~4~A and 4~B, respectively.
The corresponding parameters are
$a_1 = (6.7 \pm 0.6 ) \times 10^6, 
b_1 = 260 \pm 100,
a_2 = (3.3 \pm 0.7) \times 10^{13},
b_2 = (2.5 \pm 6.6) \times 10^9,
c_2 = (1.9 \pm 0.7) \times 10^6$.
The grey shaded region in these figures indicates the mean square fluctuation (one standard deviation) around the average. The average phase estimation uncertainty is shown as the solid orange line in Fig.~4~C. The shaded region in this figure is the mean square fluctuation of the phase uncertainty around the average value. 

For arbitrary distributions of twin Fock states, the phase uncertainty calculated according to Eq.~(\ref{ErrProp}) is not optimal for all the values of $\theta$. Quite generally, the optimal phase estimation uncertainty obtained by measuring the number of particles at the output port of the interferometer is provided by the Cramer-Rao (CR) bound 
\be \label{CR}
\Delta \theta_{\mathrm{CR}} \equiv \frac{1}{\sqrt{n\,\mathcal{F}(\theta)}}, 
\ee
where 
\be
\mathcal{F}(\theta) = \sum_{N_{+1},N_{-1}} \frac{1}{P(N_{+1},N_{-1} \vert \theta)} 
\bigg( \frac{\ud}{\ud \theta} P(N_{+1},N_{-1} \vert \theta) \bigg)^2
\ee
is the Fisher information and $P(N_{+1},N_{-1} \vert \theta)$ are the conditional probabilities to measure $N_{+1}$ and $N_{-1}$, given $\theta$. Although we do not have enough experimental data to estimate these probabilities, we can cal\-cu\-late the Fisher information within a Gaussian de\-tec\-tion noise approximation which, as shown in Fig.~2~D, is a good model of the main source of noise in our apparatus. We take, as conditional probabilities, the convolution of the ideal 
$P_{\id}(N_{+1},N_{-1} \vert \theta) 
\equiv \langle N_{+1}, N_{-1} \vert e^{-i \theta  J_x} \rho_{\mathrm{inp}} e^{+i \theta  J_x} \vert N_{+1}, N_{-1} \rangle$
\cite{Rowe2001} with Gaussian distributions of width $\sigma_\text{dn}$:
$P_{\mathrm{gdn}}(N_{+1},N_{-1} \vert \theta) \propto \sum_{N'_{+1},N'_{-1}} 
e^{-(N_{+1} - N_{+1}')^2/8\sigma_\text{dn}^2} \, 
e^{-(N_{-1} - N_{-1}')^2/8\sigma_\text{dn}^2} \times$
$P_{\id}(N_{+1},N_{-1} \vert \theta)$. 
This approximation is consistent with Eqs.~(\ref{gJz2}) and (\ref{gDJz2}).

In the ideal case ($\sigma_\text{dn}=0$), the CR bound is 
\be \label{CR_opt}
\Delta \theta_{\mathrm{CR}} = \frac{\sqrt{2}}{\sqrt{n} \, \sqrt{\langle N_{\mathrm{tot}}^2 \rangle + 
2 \langle N_{\mathrm{tot}} \rangle}},
\ee
independently of the phase shift. For arbitrary small, but finite, detection noise, $\sigma_\text{dn}>0$, the CR bound diverges at $\theta=0$: In this case, the probabilities $P(N_{+1},N_{-1} \vert \theta)$ are dominated by the detection noise and the derivative $\ud P(N_{+1},N_{-1} \vert \theta)/\ud \theta$ vanishes at $\theta=0$. For $\sigma_\text{dn}=20$, the Cramer-Rao bound Eq.~(\ref{CR}) is shown as the dotted orange line in Fig.~4~C. A comparison between the phase estimation uncertainty calculated with Eq.~(\ref{ErrProp}) and the Cramer-Rao bound with Eq.~(\ref{CR}) is shown in Fig.~\ref{FigSup} for different values of $\sigma_\text{dn}$. It is interesting to notice that, for small values of $\theta$, Eq.~(\ref{ErrProp}) follows the CR bound, thus reaching the optimal uncertainty. The difference between Eq.~(\ref{ErrProp}) and the CR bound increases with $\theta$ and, in particular, is large for small detection noise. This illustrates the possibility to further increase the sensitivity of our apparatus by decreasing the detection noise and considering a more accurate phase estimation protocol, for instance, by taking higher moments of the distribution of the population imbalance into account.
%Our current dominant source of noise is due to finite efficiency detection. Other sources of noise arise from the residual two-body interatomic interaction and atom losses due to three body recombinations~\cite{Fattori2008}.
Note that the main contribution to the detection noise is due to the electron shot noise in the camera pixels. We estimate that a reduction of the size of the atomic cloud by a factor of 2 (i.e. using shorter expansion time or Feshbach resonances~\cite{Fattori2008}) and a camera with a quantum efficiency of $50 \%$ should allow for an improved detection noise $\sigma_{\textrm{dn}} \approx 5$. 

\bibliography{Luecke}
\bibliographystyle{Luecke}
%\bibliography{klempt}

\end{document}